\documentclass[aps,prl,preprint,tightenlines,superscriptaddress,showpacs,byrevtex]{revtex4}
\usepackage{graphicx} 
\usepackage{dcolumn}  

\begin{document}

\vspace*{-3\baselineskip}
\resizebox{!}{3cm}{\includegraphics{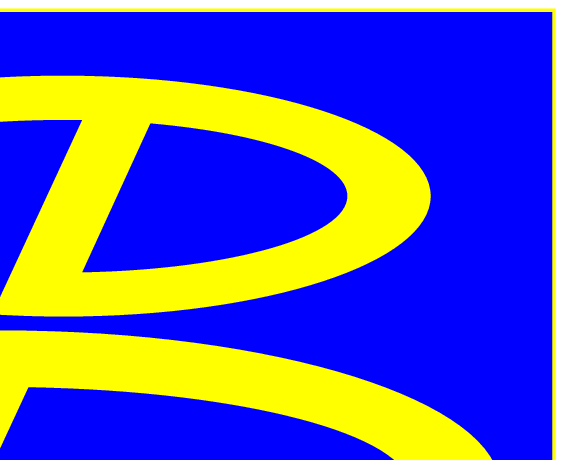}}

\preprint{\vbox{ \hbox{   }
    \hbox{Belle Preprint 2004-4}
    \hbox{KEK Preprint 2003-126}
}}

\title{\boldmath Evidence for $B^+\rightarrow\omega l^+ \nu$}

\affiliation{Budker Institute of Nuclear Physics, Novosibirsk}
\affiliation{Chiba University, Chiba}
\affiliation{University of Cincinnati, Cincinnati, Ohio 45221}
\affiliation{University of Frankfurt, Frankfurt}
\affiliation{University of Hawaii, Honolulu, Hawaii 96822}
\affiliation{High Energy Accelerator Research Organization (KEK), Tsukuba}
\affiliation{Hiroshima Institute of Technology, Hiroshima}
\affiliation{Institute of High Energy Physics, Chinese Academy of Sciences, Beijing}
\affiliation{Institute of High Energy Physics, Vienna}
\affiliation{Institute for Theoretical and Experimental Physics, Moscow}
\affiliation{J. Stefan Institute, Ljubljana}
\affiliation{Kanagawa University, Yokohama}
\affiliation{Korea University, Seoul}
\affiliation{Kyungpook National University, Taegu}
\affiliation{Swiss Federal Institute of Technology of Lausanne, EPFL, Lausanne}
\affiliation{University of Ljubljana, Ljubljana}
\affiliation{University of Maribor, Maribor}
\affiliation{University of Melbourne, Victoria}
\affiliation{Nagoya University, Nagoya}
\affiliation{National United University, Miao Li}
\affiliation{Department of Physics, National Taiwan University, Taipei}
\affiliation{H. Niewodniczanski Institute of Nuclear Physics, Krakow}
\affiliation{Nihon Dental College, Niigata}
\affiliation{Niigata University, Niigata}
\affiliation{Osaka City University, Osaka}
\affiliation{Osaka University, Osaka}
\affiliation{Panjab University, Chandigarh}
\affiliation{Peking University, Beijing}
\affiliation{Princeton University, Princeton, New Jersey 08545}
\affiliation{RIKEN BNL Research Center, Upton, New York 11973}
\affiliation{University of Science and Technology of China, Hefei}
\affiliation{Seoul National University, Seoul}
\affiliation{Sungkyunkwan University, Suwon}
\affiliation{University of Sydney, Sydney NSW}
\affiliation{Tata Institute of Fundamental Research, Bombay}
\affiliation{Toho University, Funabashi}
\affiliation{Tohoku Gakuin University, Tagajo}
\affiliation{Tohoku University, Sendai}
\affiliation{Department of Physics, University of Tokyo, Tokyo}
\affiliation{Tokyo Institute of Technology, Tokyo}
\affiliation{Tokyo Metropolitan University, Tokyo}
\affiliation{Tokyo University of Agriculture and Technology, Tokyo}
\affiliation{University of Tsukuba, Tsukuba}
\affiliation{Utkal University, Bhubaneswer}
\affiliation{Virginia Polytechnic Institute and State University, Blacksburg, Virginia 24061}
\affiliation{Yokkaichi University, Yokkaichi}
\affiliation{Yonsei University, Seoul}
  \author{C.~Schwanda}\affiliation{Institute of High Energy Physics, Vienna} 
  \author{K.~Abe}\affiliation{High Energy Accelerator Research Organization (KEK), Tsukuba} 
  \author{K.~Abe}\affiliation{Tohoku Gakuin University, Tagajo} 
  \author{T.~Abe}\affiliation{High Energy Accelerator Research Organization (KEK), Tsukuba} 
  \author{I.~Adachi}\affiliation{High Energy Accelerator Research Organization (KEK), Tsukuba} 
  \author{H.~Aihara}\affiliation{Department of Physics, University of Tokyo, Tokyo} 
  \author{M.~Akatsu}\affiliation{Nagoya University, Nagoya} 
  \author{Y.~Asano}\affiliation{University of Tsukuba, Tsukuba} 
  \author{T.~Aushev}\affiliation{Institute for Theoretical and Experimental Physics, Moscow} 
  \author{S.~Bahinipati}\affiliation{University of Cincinnati, Cincinnati, Ohio 45221} 
% ...
\collaboration{The Belle Collaboration}

\begin{abstract}
We have searched for the decay~$B^+\rightarrow\omega l^+\nu$ ($l=e$ or
$\mu$) in 78~fb$^{-1}$ of $\Upsilon(4S)$~data (85~million $B\bar
B$~events) accumulated with the Belle detector. The final state is
fully reconstructed using the $\omega$~decay into $\pi^+\pi^-\pi^0$,
combined with
detector hermeticity to estimate the neutrino momentum. A signal
of $414\pm 125$~events is found in the data, corresponding to a
branching fraction of $(1.3\pm 0.4\pm 0.2\pm 0.3)\times 10^{-4}$,
where the first two errors are statistical and systematic,
respectively. The third error reflects the estimated form-factor
uncertainty.
\end{abstract}


\maketitle


The magnitude of $V_{ub}$ plays an important role in probing the
unitarity of the Cabibbo-Kobayashi-Maskawa (CKM)
matrix~\cite{ref:1}. The cleanest way to constrain this quantity is
either by measuring the decay~$B\rightarrow X_ul\nu$~\cite{ref:2}
inclusively, or by reconstructing one of its exclusive sub-modes. As
to the latter, the decay modes $B\rightarrow\pi l\nu$ and
$B\rightarrow\rho l\nu$ have already been
observed~\cite{ref:4,ref:4b}. In this letter, we present a study of
the decay~$B^+\rightarrow\omega l^+\nu$~\cite{ref:3} which has not
been measured so far~\cite{ref:13,ref:14}. Using three different
form-factor calculations, ISGW2~\cite{ref:6}, UKQCD~\cite{ref:7} and
LCSR~\cite{ref:8}, we extrapolate the decay rates to the full range of
lepton momentum and measure the branching fraction of this decay.

The analysis is based on the data recorded with the Belle
detector~\cite{ref:5} at the asymmetric $e^+e^-$~collider
KEKB~\cite{ref:5a} operating at the center-of-mass (c.m.) energy of
the $\Upsilon(4S)$~resonance. KEKB consists of a low energy ring (LER)
of 3.5 GeV positrons and a high energy ring (HER) of 8 GeV
electrons. The $\Upsilon(4S)$~dataset used for this study corresponds
to an integrated luminosity of 78.1~fb$^{-1}$ and contains $(85.0\pm
0.5)\times 10^6$ $B\bar B$~events. In addition, 8.8~fb$^{-1}$ of data
taken at 60~MeV below the resonance are used to study the continuum
(non-$B\bar B$) background.

The Belle detector is a large-solid-angle magnetic spectrometer
consisting of a three-layer silicon vertex detector (SVD), a 50-layer
central drift chamber (CDC), an array of aerogel threshold
\v{C}erenkov counters (ACC), a barrel-like arrangement of
time-of-flight scintillation counters (TOF), and an electromagnetic
calorimeter comprised of CsI(Tl) crystals (ECL) located inside a
super-conducting solenoid coil that provides a 1.5~T magnetic
field. The responses of the ECL, CDC ($dE/dx$) and ACC detectors are
combined to provide clean electron identification. Muons are
identified in the instrumented iron flux-return (KLM) located outside
of the coil. Charged hadron identification relies on the information
from the CDC, ACC and TOF~sub-detectors.

Full detector simulation based on GEANT~\cite{ref:5b} is applied to
Monte Carlo simulated events. This analysis uses background Monte
Carlo samples equivalent to about three times the integrated
luminosity. The decay $B\rightarrow D^*l\nu$ is simulated using a
HQET-based model~\cite{ref:5c}. The ISGW2~model is used
for the decays $B\rightarrow Dl\nu$ and $B\rightarrow D^{**}l\nu$. The
modes $B\rightarrow D^{(*)}\pi l\nu$ are simulated according to the
Goity-Roberts model~\cite{ref:6b}. The ISGW2 and the De Fazio-Neubert
model~\cite{ref:8b} are used to model the cross-feed from
other decays~$B\rightarrow X_ul\nu$.

Events passing the hadronic selection~\cite{ref:8c} are required to
contain a single lepton (electron or muon) with a c.m.\
momentum~$p^*_l$~\cite{ref:8d} between 1.8 and 2.7~GeV/$c$. In this
momentum range, electrons (muons) are selected with an efficiency of
92\% (89\%) and the probability to misidentify a pion as an electron
(a muon) is 0.25\% (1.4\%)~\cite{ref:8e,ref:8f}.

The missing four-momentum is calculated,
\begin{eqnarray}
  \vec p_\mathrm{miss} & = & \vec p_\mathrm{HER}+\vec
  p_\mathrm{LER}-\sum_i\vec p_i~, \nonumber \\
  E_\mathrm{miss} & = & E_\mathrm{HER}+E_\mathrm{LER}-\sum_i E_i~,
\end{eqnarray}
where the sums run over all reconstructed charged tracks (assumed to
have the pion mass) and photons, and the labels HER and LER refer to
the two colliding beams. To reject events in which the missing
momentum misrepresents the neutrino momentum, the
following requirements are applied. The total event charge must be close
to neutral: $|Q_\mathrm{tot}|<3e$; the polar angle of the missing momentum
(with respect to the beam direction) is required to lie within the
ECL~acceptance: $17^\circ<\theta_\mathrm{miss}<150^\circ$; and the
missing mass squared, $m^2_\mathrm{miss}=E^2_\mathrm{miss}-\vec
p^2_\mathrm{miss}$, is required to be zero within about $\pm 3$
standard deviations: $|m^2_\mathrm{miss}|<3$~GeV$^2$/$c^4$.

For generic $B\rightarrow X_ul\nu$~events, the efficiency after
applying these requirements is 11\%, and the resolution in
the magnitude of the missing momentum is around 140~MeV/$c$. As the
missing energy resolution is worse than the missing momentum
resolution, the neutrino four-momentum is taken to be $(|\vec
p_\mathrm{miss}|,\vec p_\mathrm{miss})$.

Pairs of photons satisfying $E_\gamma>30$~MeV,
$p^*_{\pi^0}>200$~MeV/$c$ and 120~MeV/$c^2<m(\gamma\gamma)<150$~MeV/$c^2$
are combined to form $\pi^0$~candidates. The
decay~$\omega\rightarrow\pi^+\pi^-\pi^0$ is reconstructed using all
possible combinations of one $\pi^0$ with two oppositely charged
tracks. Combinations with a charged track identified as a kaon are
rejected, and the following requirements are imposed:
$p^*_\omega>300$~MeV/$c$,
703~MeV/$c^2<m(\pi^+\pi^-\pi^0)<863$~MeV/$c^2$. The Dalitz amplitude,
$A\propto |\vec p_{\pi^+}\times \vec p_{\pi^-}|$, is required to be
larger than half of its maximum value.

The lepton in the event is combined with the $\omega$~candidate
and the neutrino. To reject combinations inconsistent with
signal decay kinematics, the requirement $|\cos\theta_{BY}|<1.1$ is
imposed, where
\begin{equation}
  \cos\theta_{BY}=\frac{2E^*_B E^*_Y-m^2_B-m^2_Y}{2p^*_Bp^*_Y}~,
\end{equation}
and $E^*_B$, $p^*_B$ and $m_B$ are fixed to
$E^*_\mathrm{beam}=\sqrt{E_\mathrm{HER}E_\mathrm{LER}}$,
$\sqrt{E^{*2}_B-m^2_B}$ and 5.279~GeV/$c^2$, respectively. The
variables~$E^*_Y$, $p^*_Y$ and $m_Y$ are the measured c.m.\ energy,
momentum and mass of the $Y=\omega+l$~system, respectively. For
well-reconstructed signal events, $\cos\theta_{BY}$ is the cosine of
the angle between the $B$ and the $Y$~system and lies between $-1$ and
$+1$ while for background the majority of events are outside this
interval.

For each $B^+\rightarrow\omega l^+\nu$~candidate, the beam-energy
constrained mass $M_\mathrm{bc}$ and $\Delta E$ are calculated,
\begin{eqnarray}
  M_\mathrm{bc} & = & \sqrt{(E^*_\mathrm{beam})^2-|\vec
  p^*_\omega+\vec p^*_l+\vec p^*_\nu |^2}~, \nonumber \\
  \Delta E & = & (E^*_\omega+E^*_l+E^*_\nu)-E^*_\mathrm{beam}~,
\end{eqnarray}
and candidates in the range $M_\mathrm{bc}>5.23$~GeV/$c^2$ and
$|\Delta E|<1.08$~GeV are selected. On average, 2.5~combinations per
event satisfy all selection criteria, and we choose the
one with the largest $\omega$~momentum in the c.m.\ frame. Monte Carlo
simulation indicates that this choice is correct in 77\% of the signal
cases.

In $B\bar B$~events, the two $B$~mesons are produced nearly
at rest, and their decay products are uniformly distributed over the
solid angle in the c.m.\ frame. Conversely, continuum events have a
jet-like topology. We exploit this property to suppress continuum
background with the following quantities (defined in the c.m.\ frame):
the ratio~$R_2$ of the second to the zeroth Fox-Wolfram
moment~\cite{ref:10} which tends to be close to zero (unity) for
spherical (jet-like) events; $\cos\theta_\mathrm{thrust}$, where
$\theta_\mathrm{thrust}$ is the angle between the thrust axis of the
$\omega l$~system and the thrust axis of the rest of the event; and a
Fisher discriminant~\cite{ref:11} that selects events with a uniform
energy distribution around the lepton direction. The input variables
to the latter are the charged and neutral energy in nine cones of
equal solid angle around the lepton momentum axis. The selection
${\mathcal L}_\mathrm{S}/({\mathcal L}_\mathrm{S}+{\mathcal
  L}_\mathrm{B})>0.9$ is applied, where ${\mathcal L}_\mathrm{S}$
(${\mathcal L}_\mathrm{B}$) is the product of the signal (background)
p.d.f.'s of these three quantities. This selection is 56\%~efficient
for signal decays and eliminates 92\% of the remaining continuum
background.

The signal yield is determined by a three-dimensional binned maximum
likelihood fit taking into account finite Monte Carlo
statistics~\cite{ref:12}. We use nine 240~MeV wide bins in $\Delta E$,
eight 20~MeV/$c^2$ wide bins of $m(\pi^+\pi^-\pi^0)$, and three
300~MeV/$c$ wide $p^*_l$ bins.  The signal resolutions in $\Delta E$ and
$m(\pi^+\pi^-\pi^0)$ are about 140~MeV and 11~MeV/$c^2$, respectively.
The backgrounds from the remaining continuum events and
from $B\bar B$~events in which the lepton is misidentified or does not
originate directly from a $B$~decay are subtracted from the raw yield
bin-by-bin. The continuum background is estimated using the
off-resonance data (scaled to the on-resonance luminosity). The
fake and non-primary lepton
backgrounds (which account for only about 2\% of the raw yield) are
determined from the simulation. The signal yield, the background from
$B\rightarrow X_ul\nu$~decays and the background from $B\rightarrow
X_cl\nu$~decays are fitted after this subtraction. In the region
defined by 763~MeV/$c^2<m(\pi^+\pi^-\pi^0)<803$~MeV/$c^2$, $|\Delta
E|<360$~MeV the signal purity is almost three times higher than the
average in the whole fitting space; the broader fit ranges in $\Delta
E$, $m(\pi^+\pi^-\pi^0)$ permit a reliable determination of the
signal and background components. The distribution shapes of the three
fit components are taken from Monte Carlo simulation. The contribution
of each component is described by a single parameter.

\begin{figure}
  \begin{center}
    \includegraphics[width=12.5cm]{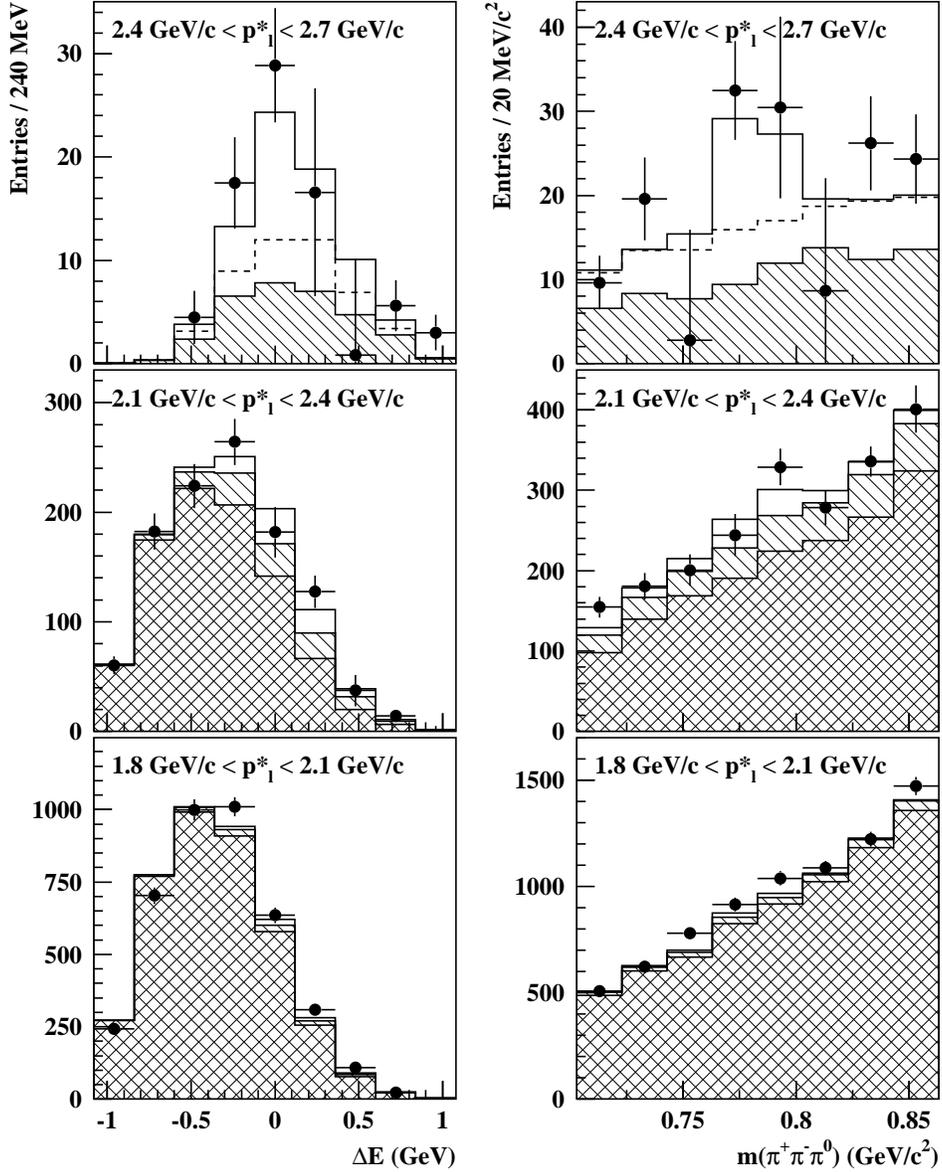}
  \end{center}
  \caption{The $\Delta E$ and $m(\pi^+\pi^-\pi^0)$ projections of the
    fit assuming ISGW2 form-factors for $B^+\rightarrow\omega l^+\nu$
    after requiring 763~MeV/$c^2<m(\pi^+\pi^-\pi^0)<803$~MeV/$c^2$ and
    $|\Delta E|<360$~MeV, respectively. The data
    points are the background subtracted yields. The open, hatched and
    doubly-hatched histograms correspond to the $B^+\rightarrow\omega
    l^+\nu$~signal, the $B\rightarrow X_ul\nu$~background and the
    $B\rightarrow X_cl\nu$~background, respectively. In the highest
    momentum bin, the contribution of signal candidates in which the
    lepton stems from a $B^+\rightarrow\omega l^+\nu$~decay but the
    $\omega$ is reconstructed improperly is shown by the dashed
    histogram.}
    \label{fig:1}
\end{figure}

Table~\ref{tab:1} and Fig.~\ref{fig:1} show the result of the fit
assuming ISGW2 form-factors for $B^+\rightarrow\omega l^+\nu$. We find $383\pm
118$~signal events with a statistical significance of 3.8~standard
deviations. The latter is defined as $\sqrt{-2\ln({\mathcal
L}_\mathrm{B}/{\mathcal L}_\mathrm{S+B})}$, where ${\mathcal
L}_\mathrm{S+B}$ (${\mathcal L}_\mathrm{B}$) refers to the maximum of
the likelihood function describing signal and background (background
only). In addition to well-reconstructed signal decays, the
signal component of the fit also includes candidates in which the lepton
stems from a $B^+\rightarrow\omega l^+\nu$~decay but the $\omega$
failed to be reconstructed properly. This sub-component is
shown separately in Fig.~\ref{fig:1}. It scales with the actual signal
and amounts to 36\% of the signal component within the
763~MeV/$c^2<m(\pi^+\pi^-\pi^0)<803$~MeV/$c^2$, $|\Delta
E|<360$~MeV region.
\begin{table}
  \caption{The result of the fit assuming ISGW2~form-factors for
  $B^+\rightarrow\omega l^+\nu$. The uncertainties quoted are
  statistical only.}

  \begin{center}
     {\small
        \begin{tabular}{cc@{\extracolsep{.5cm}}ccc}
        \hline \hline
        \rule{0pt}{2.7ex} & \multicolumn{4}{c}{$p^*_l$ range (GeV/$c$):}\\
        \rule[-1.3ex]{0pt}{1.3ex} & 1.8 -- 2.1 & 2.1 -- 2.4 & 2.4 --
        2.7 & 1.8 -- 2.7\\
        \hline
        \rule{0pt}{2.7ex}raw yield & 16,777 & 4639 & 326 & 21,742\\
        continuum & $234\pm 45$ & $294\pm 50$ & $78\pm 26$ & $606\pm 72$\\
        other bkgrds.\ & $211\pm 9$ & $216\pm 9$ & $28\pm 3$ & $455\pm 13$\\
        \rule[-1.3ex]{0pt}{1.3ex}{\bf subtr.\ yield} & $16,332\pm 46$ &
        $4129\pm 51$ & $220\pm 26$ & $20,681\pm 74$\\
        \hline
        \rule{0pt}{2.7ex}$B^+\rightarrow\omega l^+\nu$ & $101\pm
        31$ & $193\pm 59$ & $89\pm 28$ & $383\pm 118$\\
        $B\rightarrow X_ul\nu$ & $339\pm 151$ & $466\pm 142$ & $125\pm19$ &
        $930\pm 312$\\
        $B\rightarrow X_cl\nu$ & $15,755\pm 238$ & $3592\pm 64$ & 0 &
        $19,348\pm 289$\\
        \rule[-1.3ex]{0pt}{1.3ex}{\bf sum} & $16,196\pm 284$ & $4251\pm
        166$ & $215\pm 34$ & $20,662\pm 442$\\
        \hline \hline
        \end{tabular} }
  \end{center}
  \label{tab:1}
\end{table}

The fit is repeated for the three form-factor models considered for
$B^+\rightarrow\omega l^+\nu$. For each model, the signal yield
$N(B^+\rightarrow\omega l^+\nu)$ is determined and the branching
fraction ${\mathcal B}(B^+\rightarrow\omega l^+\nu)$ is calculated
according to the relation $N(B^+\rightarrow\omega
l^+\nu)=N(B^+)\times{\mathcal B}(B^+\rightarrow\omega l^+\nu)\times
{\mathcal
B}(\omega\rightarrow\pi^+\pi^-\pi^0)\times(\epsilon_e+\epsilon_\mu)$,
where $N(B^+)$ is the total number of charged $B$~mesons in the
data (assumed to be equal to the number of $B\bar B$~events), ${\mathcal
  B}(\omega\rightarrow\pi^+\pi^-\pi^0)=(89.1\pm 0.7)\%$~\cite{ref:9}
and $\epsilon_e$ ($\epsilon_\mu$) is the model-dependent selection
efficiency for $\omega e\nu$ ($\omega\mu\nu$) signal candidates. The
results are presented in Table~\ref{tab:2}. Averaging
the central values and the statistical uncertainties over the three
models (giving equal weight to each), a branching fraction of $(1.3\pm
0.4)\times 10^{-4}$ is obtained. The spread around this average value
amounts to $0.3\times 10^{-4}$, and provides an estimate of
the form-factor model uncertainty.
\begin{table}
  \caption{The fitted signal yield, the selection efficiency for signal
    candidates, the branching fraction and the goodness of fit
    (estimated by the $\chi^2$ divided by the number of degrees of
    freedom). For each fit (using a given form-factor model), the
    error quoted on the signal yield and the branching fraction is
    statistical only. For the average, the first error is statistical,
    and the second is the spread around the central value.} 
  \begin{center}
    \begin{tabular}{c@{\extracolsep{.5cm}}cccc}
      \hline \hline
      \rule[-1.3ex]{0pt}{4ex}form-factor model &
      $N(B^+\rightarrow\omega l^+\nu)$ & $\epsilon_e+\epsilon_\mu$ &
      ${\mathcal B}(B^+\rightarrow\omega l^+\nu)/10^{-4}$ &
      $\chi^2/\mathrm{ndf}$\\
      \hline
      \rule{0pt}{2.7ex}ISGW2~\cite{ref:6} & $383\pm 118$ & 5.0\% &
      $1.0\pm 0.3$ & 1.05\\
      UKQCD~\cite{ref:7} & $384\pm 116$ & 4.2\% & $1.2\pm 0.4$ & 1.08\\
      \rule[-1.3ex]{0pt}{1.3ex}LCSR~\cite{ref:8} & $473\pm 141$ &
      3.8\% & $1.7\pm 0.5$ & 1.04\\
      \hline
      \rule[-1.3ex]{0pt}{4ex}average & $414\pm 125\pm 42$ & & $1.3\pm
      0.4\pm 0.3$ & \\
      \hline \hline
    \end{tabular}
  \end{center}
  \label{tab:2}
\end{table}

The experimental systematic error is 18.1\% of the
branching fraction (Table~\ref{tab:3}), or $0.2\times 10^{-4}$ in
absolute. The largest contribution is the uncertainty in the
$X_ul\nu$~cross-feed. It is estimated by separately varying the
fraction of $B\rightarrow\pi l\nu$ and $B\rightarrow\rho l\nu$~decays
(which are expected to dominate in the high $p^*_l$~region) within their
respective experimental uncertainties~\cite{ref:9}. The relative
fractions of charged and neutral modes are constrained using isospin
symmetry. For the $B\rightarrow\rho l\nu$~mode, we also consider the
form-factor model dependence of the
cross-feed~\cite{ref:6,ref:7,ref:8}. To estimate the uncertainy in the
cross-feed from other $B\rightarrow X_ul\nu$~decays, the fit is
repeated modeling this component once with ISGW2, and once with the De
Fazio-Neubert model. Half of the difference between these two cases is
assigned as a systematic uncertainty. The next-to-largest component is
the uncertainty in the neutrino reconstruction, track finding and cluster
finding efficiency. While the latter two are uncorrelated, they are
treated as fully correlated with the former. The $X_cl\nu$~cross-feed
uncertainty is estimated by varying the fractions of $B\rightarrow
D^*l\nu$, $B\rightarrow Dl\nu$ and $B\rightarrow D^{**}/D^{(*)}\pi
l\nu$ in $B\rightarrow X_cl\nu$ within $\pm 10\%$, $\pm 10\%$ and $\pm
30\%$, respectively, and summing the individual variations in
quadrature.
\begin{table}
  \caption{Contributions to the systematic uncertainty. The size of
    each contribution is given as percentage of the branching
    fraction.}
  \begin{center}
    \begin{tabular}{l@{\extracolsep{1cm}}c}
      \hline \hline
      \rule[-1.3ex]{0pt}{4ex} & $\Delta{\mathcal B}/{\mathcal B}$\\
      \hline
      \rule{0pt}{2.7ex}$B\rightarrow\pi l\nu$ cross-feed & 2.2\%\\
      $B\rightarrow\rho l\nu$ cross-feed & 14.8\%\\
      other $B\rightarrow X_ul\nu$ cross-feed & 1.4\%\\
      \rule[-1.3ex]{0pt}{1.3ex}{\bf (sum)} & 15.1\%\\
      \hline
      \rule{0pt}{2.7ex}neutrino reconstruction & 4\%\\
      charged track finding ($l,\pi^+,\pi^-$) & 3\%\\
      cluster finding ($\pi^0$) & 4\%\\
      \rule[-1.3ex]{0pt}{1.3ex}{\bf (sum)} & 9\%\\
      \hline
      \rule{0pt}{2.7ex}$X_cl\nu$ cross-feed & 2.7\%\\
      lepton identification & 3.0\%\\
      number of $B\bar B$ & 0.6\%\\
      \rule[-1.3ex]{0pt}{1.3ex}$\omega\rightarrow\pi^+\pi^-\pi^0$
      branching fraction~\cite{ref:9} & 0.8\%\\
      \hline
      \rule[-1.3ex]{0pt}{4ex}{\bf total systematic uncertainty} &
      18.1\%\\
      \hline \hline
    \end{tabular}
  \end{center}
  \label{tab:3}
\end{table}

In summary, we have measured the $B^+\rightarrow\omega l^+\nu$
branching fraction to be $(1.3\pm 0.4(stat)\pm 0.2(syst)\pm
0.3(model))\times 10^{-4}$, based on $414\pm 125$~signal events. This
is the first evidence for this decay. Assuming the quark model
relation $\Gamma(B^0\rightarrow\rho^-l^+\nu)=2\Gamma(B^+\rightarrow\omega
l^+\nu)$, our measurement agrees with measurements of the
decay~$B^0\rightarrow\rho^-l^+\nu$~\cite{ref:4,ref:4b}.

We thank the KEKB group for the excellent operation of the
accelerator, the KEK Cryogenics group for the efficient operation
of the solenoid, and the KEK computer group and the NII for
valuable computing and Super-SINET network support.  We acknowledge
support from MEXT and JSPS (Japan); ARC and DEST (Australia); NSFC (contract
No.~10175071, China); DST (India); the BK21 program of MOEHRD and the
CHEP SRC program of KOSEF (Korea); KBN (contract No.~2P03B 01324,
Poland); MIST (Russia); MESS (Slovenia); NSC and MOE (Taiwan); and DOE
(USA).


\begin{thebibliography}{99}
\bibitem{ref:3} Throughout this letter, the inclusion of the charge
  conjugate mode is implied.
\bibitem{ref:8d} Throughout this letter, quantities calculated in the
  c.m.\ frame are denoted by an asterisk.
\end{thebibliography}
\end{document}